\documentclass[a4paper]{mem}
\usepackage{natbib}
\usepackage{epsfig}
\usepackage{graphicx}
\usepackage[a4paper]{hyperref}
\newcommand{\chandra}{\emph{Chandra}}
\newcommand{\xmm}{XMM-\emph{Newton}}
\begin{document}
\title{Accretion, fluorescent X-ray emission and flaring magnetic
structures in YSOs}

\author{F. Favata 
}
\offprints{F. Favata}
\mail{Fabio.Favata@rssd.esa.int}

\institute{Astrophysics Division of ESA-ESTEC, P.O. Box 299, 2200 AG
  Noordwijk, The Netherlands \email{Fabio.Favata@rssd.esa.int} }

\abstract{I present some recent developments on high-energy
   phenomena in YSOs, concentrating on the new
   evidence for accretion-induced X-ray emission in YSOs, 
   for Fe 6.4 keV fluorescent emission from the disks of YSOs and for
   very long magnetic structures responsible for intense X-ray flares,
   likely connecting the star and the circumstellar disk.
   }
   \authorrunning{F. Favata}
   \titlerunning{Accretion, fluorescence and magnetic
structures in YSOs} 
   \maketitle
%

\section{Introduction}


YSOs were identified as copious sources of X-ray emission in the early
days of imaging X-ray astronomy, immediately raising a number of
long-debated questions. Significant progress has been achieved in the
last couple of years on a number of them, thanks to the availability
of new observational material. I will discuss here three of these
questions. 

X-ray emission starts to become present in Class I objects, with no
confirmed detection of X-ray emission from Class 0 sources. Class II
(CTTS) and Class III (WTTS) objects are almost always detected as
strong X-ray sources. The possible difference in the X-ray behavior
and in the emission mechanisms between CTTSs (with active, accreting
disks) and WTTSs (with no ongoing accretion, and thus less disturbed
photospheres) has been an open topic until very recently. In
particular the role of accretion has been often debated, but with very
little observational evidence.  The recent availability of
high-resolution X-ray spectra of CTTS is finally allowing to address
the issue, and is showing that accretion is indeed causing a part of
the X-ray emission observed in CTTS, with much interesting physics
going on.

Once it became evident that YSOs are strong and persistent X-ray
sources, the question immediately arose how and to which extent they
affect the circumstellar environment, and in particular the accretion
disk and its proto-planetary environment. The details of the geometry
(e.g. the relative position of the X-ray sources and of the disk
material, the degree to which the X-rays penetrate the disk, which
depends on the angle of incidence, etc.) are critical, but direct
evidence for the amount of interaction between stellar X-rays and the
disk has only been available recently, thanks to the \chandra\ and
\xmm\ observations of fluorescent X-rays from the disks of YSOs.

The geometry of the magnetic field on and around YSOs has also been a
topic of debate. Photospheric fields have been measured, and range
into the kG domain, but how these extend further from the stellar
surface, and how do they link to the disk structure, is not
obvious. The magnetospheric model of accretion postulates flux tubes
connecting the disk's inner boundary and the photosphere, but whether
such structures are also the seat of coronal plasma is not known. The
evidence available to date using the analysis of flare decay on older
stars invariably points to compact coronal structures, extending at
most a fraction of the stellar radius from the photosphere. New long
X-ray observations of a large number of YSOs in the ONC (the COUP
project, presented elsewhere in this book) have allowed to extend this
type of analysis to YSOs.

\section{Accretion and X-ray emission}

The generally accepted model for accretion from the disk into YSOs
(the magnetospheric model) envisions plasma being channeled into
magnetic flux tubes and ramming onto the star at essentially free-fall
speed. For normal YSO parameters (mass and radius) this implies shock
temperatures of up to a few MK, and thus possible shock-related X-ray
emission. However, the X-ray emission from the majority of CTTS is
dominated by much hotter plasma (up to some tens of MK), which cannot
be directly produced by accretion shocks. The first \chandra\ 
high-resolution spectra of a CTTS, TW Hya, therefore came as somewhat
of a surprise, in that its density-sensitive O\,{\sc vii} He-like
triplet (as well as the Ne triplet) show very low values of the ratio
between the forbidden and intercombination lines ($R = f/i$)
\citep{kas2002}, a result confirmed by the \xmm\ observation
\citep{ss2004}. $R$ is a density-sensitive line ratio, and the low
values observed in TW Hya are compatible with very high coronal
densities, $n_e \simeq 10^{13}$ cm$^{-3}$ for the plasma at
temperatures 1--3 MK (at which the O\,{\sc vii} triplet is formed).
Such densities are never observed in older active stars, where the
O\,{\sc vii} $R$ ratio typically implies densities of at most a few
times $10^{11}$ cm$^{-3}$.

\begin{figure}
  \centering
  \includegraphics[width=5cm]{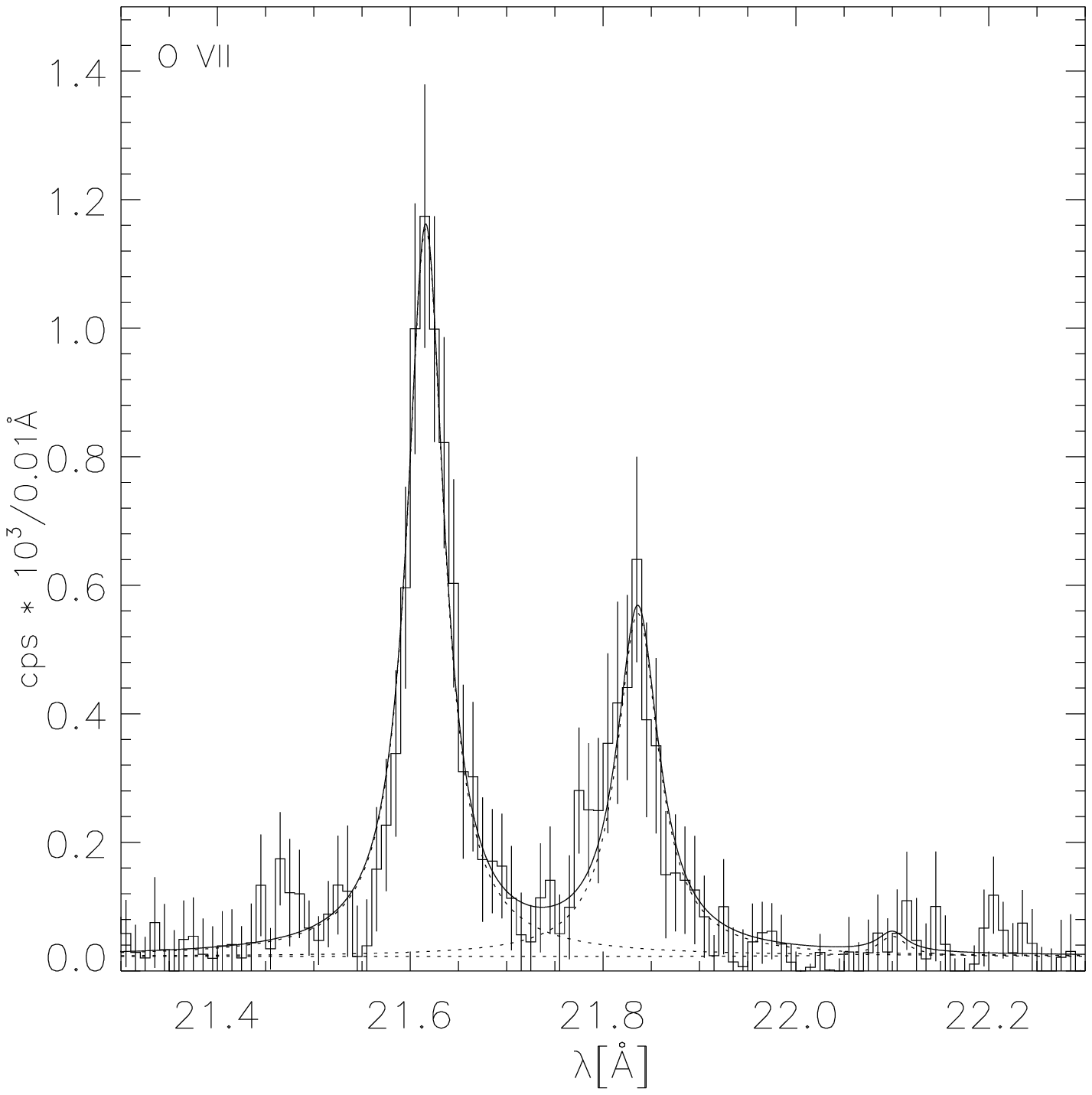}\vspace{17ex}
  \epsfig{width=5cm, angle=180, file=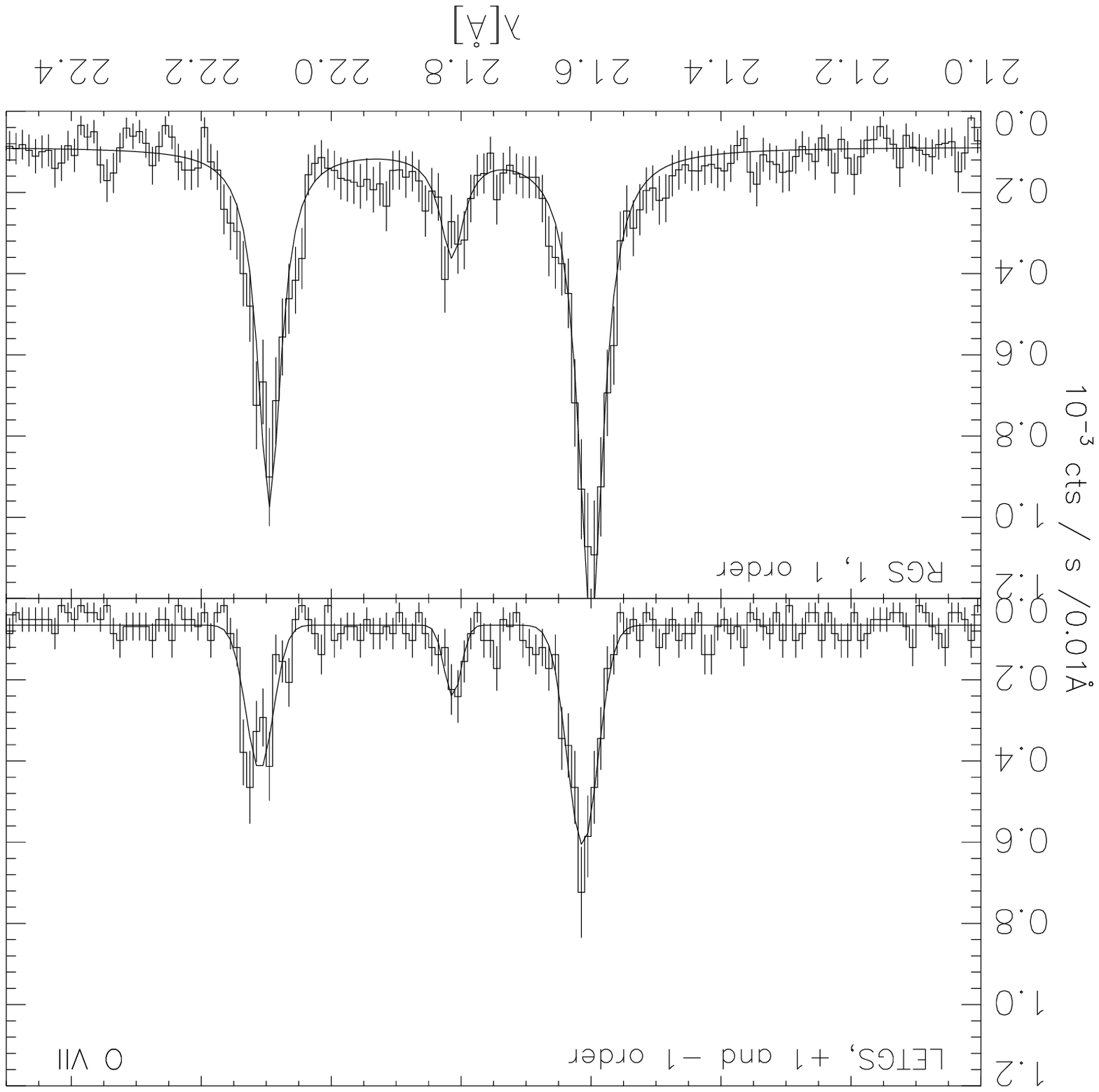, bbllx=10, bblly=10, bbury=283,
    bbllx=500, clip=} 
  \caption{The O\,{\sc vii} triplet in TW Hya as observed by \xmm,
    compared with the same spectral region observed in a normal
    coronal source (YY Gem, bottom panel). The absence of the
    forbidden line in TW Hya is evident.  Figure courtesy of B.
    Stelzer. }
  \label{fig:tw}
\end{figure}

TW Hya is also peculiar in its low-resolution X-ray spectrum, which
shows no significant amounts of plasma hotter than $\simeq 3$ MK, so
that the bulk of TW Hya's X-ray emission is compatible with being
produced in an accretion shock. But, while high densities are expected
in a shock, they are not the only possible cause of low O\,{\sc vii}
triplet $R$ values: high UV fluxes can radiatively depopulate the
atomic level associated with the $f$ line and thus result in low $R$
independently from the density (low $R$ values are in fact normally
observed in early type stars, where the UV pumping is clearly the
cause). While \citet{kas2002} and \citet{ss2004} both interpreted the
observed $R$ value as due to high density, \citet{dra2005} has shown
that this interpretation has a problem: due to the density structure
of the shock, to reach $n_e = 10^{13}$ cm$^{-3}$ in the X-ray emitting
region the shock should be buried below such a high column density of
material as to totally absorb the emitted (soft) X-rays. On the other
hand, if indeed the X-ray emission is shock-emitted, its association
with the accretion hot spot (which has typical temperatures of
8000--10\,000 K) would embed the X-ray emitting plasma in a strong UV
flux, which would induce a low $R$ value. As the two effects cannot be
distinguished, the actual density of the plasma cannot be determined
directly; nevertheless, given the very small filling factor ($f \simeq
1$\%) of the accreting spot, to be affected by the UV flux the plasma
must be located very close to the accretion spot. Therefore,
\emph{independently from the relative contribution of UV flux and high
  density}, the low observed $R$ value points to a strict association
between the accretion spot and the X-ray emission.

TW Hya has been until very recently the only CTTS observed in X-rays
at high spectral resolution, raising the question of whether the low
$R$ ratios and therefore the associated physics are a common
phenomenology in CTTS or whether TW Hya itself is a peculiar object.
Another CTTS, BP Tau, has been observed at high resolution during the
summer by \xmm, and the result \citep{sch2005} is that also in BP Tau
the O\,{\sc vii} $R$ ratio is very low, again pointing at a close
association between the 1--3 MK plasma and the accretion spot.
However, unlikely TW Hya, BP Tau has significant amounts of hotter
plasma, up to $\simeq 30$ MK, which cannot be produced in the
accretion spot, as not enough gravitational energy is available. The
hotter plasma must then be produced in some form of coronal process,
i.e. from magnetically confined -- and heated -- plasma. The magnetic
confinement in other CTTS is made evident by the presence of flares
with a decay typical of flares in older stars, and implying
confinement. Therefore, it appears that, in CTTS, both accretion
produced X-ray and coronal X-rays can coexist, and that production of
significant soft X-ray flux in the accretion shock in YSOs possibly is
a common, general phenomenon. As discussed in Sect.~\ref{sec:flare},
the magnetic structures responsible for the confinement of the hotter
plasma, may also be confining the accreting material.

\section{X-ray fluorescence in accretion disks around YSOs}

The material in the accretion disk in a YSO will be exposed to stellar
X-rays (whether coronal or accretion produced); exposure to
high-energy radiation has in fact been invoked as the mechanism
explaining e.g. the isotopic anomalies observed in some types of
meteorites, as well as the recent detection of calcite around some
YSOs \citep{chi2004}; in terrestrial condition the formation of
calcite requires liquid water, and it is therefore difficult to
picture how calcite could form in the cold conditions of an accretion
disk; temporary thawing in ice particles induced by absorption of
energetic photons could however lead to calcite formation
\citep{chi2004}.

Until recently, the evidence for the interaction of stellar X-rays
with the circumstellar material has been indirect. Direct evidence for
interaction of X-rays with disk material is now present in the form of
detection of fluorescent X-ray emission from ``cold'' material. The
X-ray spectrum of hot plasma ($T \ge 30$ MK) is characterized by
emission from a complex of lines around 6.7 keV, originating from
Fe\,{\sc xvv}, which is typically very evident (as an unresolved line)
in low resolution X-ray spectra of YSOs. If cold material ($T < 1$ MK)
is irradiated with energetic X-rays ($E \ge 7.11$ keV), neutral Fe
will be photoexcited, and will produce X-rays in a number of lines
centered at around 6.4 keV. The equivalent width of the 6.4 keV line
(also unresolved in low-resolution spectra) depends on a number of
factors, such as the intensity and spectrum of the incident radiation,
the fluorescent efficiency, and the amount of fluorescing material and
its geometry.

Fluorescent Fe emission at 6.4 keV was first detected during an
intense flare in a Class I source in $\rho$ Oph (YLW 16A) by
\citet{ima2001}, and recently the same phenomenon has been observed in
a number of YSOs in the ONC \citep{tsu2004}, using the very long COUP
\chandra\ observation. In all these cases, the fluorescent emission
was associated with intense flares, which produced the high-energy
photons needed to excite the fluorescent emission. The low quiescent
X-ray flux from the ONC sources (given their distance) does not allow
to determine if the fluorescent X-ray emission is only associated with
intense flares (and thus is a sporadic, transient phenomenon) or
whether is a persistent feature in these stars.

Also very recently, a study of the \chandra\ and \xmm\ observations of
another Class I source in $\rho$ Oph, Elias 29 \citep{fav2005}, has
shown intense Fe 6.4 keV fluorescent line emission both during the
quiescent phases of its coronal emission and during a moderate
intensity flare.

   \begin{figure}
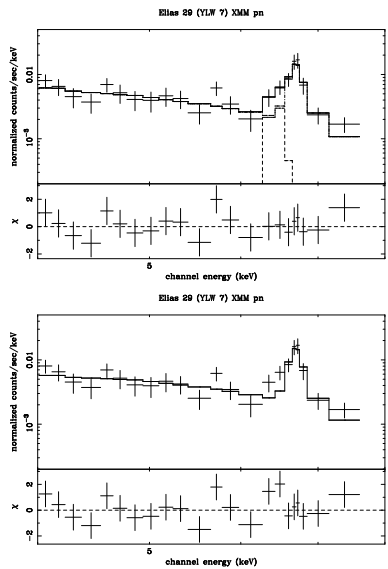

   \centering
   \includegraphics[angle=-90,width=5cm]{el29pnspw64.ps}\vspace{1ex}
   \includegraphics[angle=-90,width=5cm]{el29pnspwo64.ps}
      \caption{The \xmm\ EPIC pn spectrum of Elias 29 (from
\citealp{fav2005}). The bottom panel shows the best-fit spectrum
assuming emission from a hot plasma, the top panel shows the
additional 6.4 keV fluorescent line from ``cold'' Fe needed to ensure a
good fit to the data.
              }
         \label{fig:el29}
\end{figure}

The equivalent width of the 6.4 keV emission observed in Elias 29 is
large ($\simeq 140$ eV, \citealp{fav2005}). This is a powerful
diagnostic of the geometry of the fluorescing material; values $\ge
100$ eV exclude, given the spectrum of the coronal X-ray emission from
the star (from which the photons exciting the fluorescence are coming)
that the emission can come from diffuse, homogeneous circumstellar
material. It also allows to exclude that the fluorescence is due to
reflection from either a photosphere or a circumstellar disk
illuminated from above. In fact, equivalent widths $\ge 100$ eV can
only be produced \citep{gf91} by a disk illuminated by a source placed
in a central hole. In addition, the disk must be seen ``face on'', as
the equivalent width rapidly decreases with disk inclination. In the
case of Elias 29, the low inclination of the disk axis relative to the
line of sight has been independently derived based on the IR spectrum
\citep{boo2002}. The centrally illuminated disk geometry is perfectly
compatible with a YSO system, in which the X-ray source is located
close to the star, and therefore only illuminates the disk from inside
the central gap. This excludes more exotic geometries, such as coronal
loops located solely on the accretion disk, but it's compatible with
the type of loops interconnecting the star and the disk discussed in
Sect.~\ref{sec:flare}.

\begin{figure*}
  \centering
  \resizebox{10cm}{!}{\includegraphics{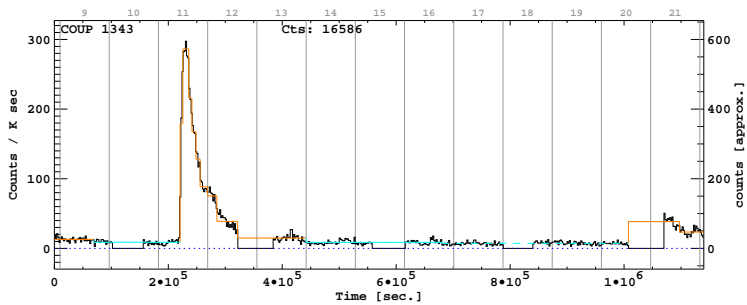}}
  \caption{The intense X-ray flaring event observed by the \chandra\
    ACIS instrument in the ONC star COUP source 1343. }
  \label{fig:flare}%
\end{figure*}

The face-on orientation is the statistically least probable one, so
that it's likely that Fe 6.4 keV fluorescent emission is a common
occurrence among YSOs, but that only a small fraction of them will
result in an observable line in the spectrum, due to the inclination
constraints. As the conditions in the disk are optically thick with
respect to the X-ray energies involved (the derivation of the
equivalent width by \citealp{gf91} is the result of a full radiation
transfer treatment), the observation of the Fe 6.4 keV fluorescence
line provides a quantitative measure of the influence of the stellar
X-ray emission onto the circumstellar disk, a necessary input to
e.g. understand its effect on the disk chemistry. A number of deep
X-ray observation campaigns are ongoing or recently proposed, which
will hopefully lead to additional detection of objects like Elias 29.

\section{Size of flarings structures}
\label{sec:flare}

The analysis of the decay of powerful stellar X-ray flares allows to
determine the size of the flaring structure, and therefore to derive a
number of important quantities, such as the density of the flaring
plasma and the strength of the confining magnetic field. In the last
years, a number of well-resolved flares have been studied on older
active stars (see \citealp{fm2003} for a review), invariably resulting
in relatively compact flaring structures, smaller than the star
itself. The COUP sample allows for the first time to extend this
analysis to a significant number of YSOs, with a number of large
flares present in the data set. Some 30 events have sufficient
statistics to allow a detailed study, and their analysis
\citep{fav2005b} reveals a broad range of physical parameters for the
flaring structures. The most notable result is that in some cases the
size of the flaring structure turns out to be much larger than
observed in older stars. One excellent example is the flare shown in
Fig.~\ref{fig:flare}: at its peak the flare outshines the quiescent
emission by a factor of $\simeq 30 \times$, and the flare lasts for
more than two days. The observed peak temperature is well above 100 MK
(the precise value cannot be determined due to the the characteristics
of the \chandra\ ACIS spectrometer, whose spectral range only allows
to put a lower limit to temperature $\ge 100$ MK), and it decays
rapidly, allowing to constrain the presence of sustained heating
during the decay. The long $1/e$ decay time, $\tau = 40$ ks, can only
be explained by the decay of a very long magnetically confined flaring
loop, with a semi-length $L \simeq 10^{12}$~cm. Shorter flaring loops
would decay more rapidly, due to the combined influence of higher
density (and thus higher emissivity) and larger conductive losses
toward the chromosphere. The resulting plasma density is $n_e \simeq
2\times 10^{10}$ cm$^{-3}$, and the minimum magnetic field necessary
to confine the plasma at the maximum temperature is 150 G. Source 1343
in the COUP sample is the best example of a very long flaring
structure, but by no means the only one, so that these structures
appear not to be exceptional in YSOs.

Such large coronal structures are not at all similar to the ones seen
in older stars, which typically have $L < R_*$. While no radius
determinations are available for COUP source 1343, assuming the
typical radius at this age for a low-mass star of a few solar radii,
one finds $L \simeq 10 R_*$. The size of the flaring loop thus is a
non-negligible fraction of an AU ($L \simeq 0.1$ AU), so that the
flaring structure extends (together with the associated magnetic
field) significantly in what would be considered ``circumstellar
space''.

Whether these large magnetic structures containing hot plasma actually
link the star with the circumstellar disk cannot be established
directly, but it certainly is a possibility. In fact, given that
solar, and by extension stellar, coronae appear to be heated by flux
braiding at the footpoints of the loops, it's very likely that the
flux tubes responsible for the magnetospheric accretion, would also be
subject to the same process (with plenty of flux shear also available
to the footpoints anchored onto the disk), so that the falling plasma
could be heated by magnetic processes. In fact, detailed modeling of
UV excesses from YSOs shows that the plasma in the flux tube
channeling the accretion flow is likely heated, producing an UV excess
above the one produced by the accretion spot.

\section{Conclusions}

Magnetically confined plasmas and the attendant high-energy processes
are a common feature of YSOs, which plays an essential role in e.g.\ 
affecting chemical processes in the circumstellar disk.  As discussed
above, the recent detection of X-ray fluorescence from circumstellar
disks shows that indeed this is likely to be a common process, and
provides a quantitative basis for the modeling of the chemical effects
of high energy radiation. In addition, the detection of very long (up
to $\simeq 0.1$ AU) flaring structure on YSOs (an unique feature, not
present among older active stars) shows that magnetic activity may
influence accretion, if the long flux tubes in which the flares are
taking place are indeed the same which channel the accretion flow.

High-energy processes are therefore not a ``minor constituent'' of star
formation, but rather one essential element, with broad reaching
implications. And, we are just beginning to unravel some very
interesting pieces of the puzzle.

\begin{acknowledgements}
  
  A significant part of the work I have presented here is the result
  of collaborative efforts. I would like to thank my colleagues and
  friends G. Micela, S. Sciortino, E. Flaccomio, F. Reale for the
  long-standing collaboration. Special thanks also go to
  J.\,H.\,M.\,M. Schmitt for allowing me to use key material (the BP
  Tau spectra) prior to their publication and to B. Stelzer for
  providing Fig.~\ref{fig:tw}.

\end{acknowledgements}

\bibliographystyle{aa}

\end{document}